\documentclass[conference]{IEEEtran}
\IEEEoverridecommandlockouts

\usepackage{cite}
\usepackage{amsmath,amssymb,amsfonts}
\usepackage{graphicx}
\usepackage[table]{xcolor}
\usepackage{booktabs}
\usepackage{array}
\usepackage{adjustbox}
\usepackage{caption}
\usepackage[hyphens]{url}
\usepackage{tcolorbox}

\definecolor{linkBlue}{RGB}{0,84,159}
\definecolor{citeTeal}{RGB}{0,128,102}
\definecolor{urlViolet}{RGB}{136,46,152}
\usepackage[colorlinks=true,linkcolor=linkBlue,citecolor=citeTeal,urlcolor=urlViolet,breaklinks]{hyperref}
\urldef{\repourl}\url{https://github.com/manoharvellala/Hardeval}

\definecolor{hePanelFrame}{RGB}{54,79,107}
\definecolor{hePanelBack}{RGB}{245,248,252}
\definecolor{heHeat}{RGB}{192,57,43}

\newtcolorbox{promptpanel}[1]{
  colback=hePanelBack, colframe=hePanelFrame,
  boxrule=0.6pt, arc=1.5pt,
  left=3pt, right=3pt, top=2pt, bottom=2pt,
  fonttitle=\bfseries\footnotesize\sffamily,
  title={#1}
}

\begin{document}

\title{HardEval: An Execution-Verified Benchmark for AI-Driven Security Hardening and Compliance Remediation\\
\thanks{The HardEval benchmark, source code, model predictions, and per-rule results are publicly available at \repourl.}%
}

\author{
    \IEEEauthorblockN{
        Manohar Vellala\IEEEauthorrefmark{1}\IEEEauthorrefmark{2},
        Soumya Banerjee\IEEEauthorrefmark{1}\IEEEauthorrefmark{3},
        Sandip Roy\IEEEauthorrefmark{3},
        Ross Gore\IEEEauthorrefmark{3},
        Sachin Shetty\IEEEauthorrefmark{1}
    }
    \IEEEauthorblockA{
        \IEEEauthorrefmark{1}Center for Secure \& Intelligent Critical Systems,
        \IEEEauthorrefmark{2}Department of Computer Science,
        \IEEEauthorrefmark{3}School of Cybersecurity\\
        Old Dominion University, Virginia, USA\\
        \{mvell001, s1banerj, sroy, rgore, sshetty\}@odu.edu
    }
}

\maketitle

\begin{abstract}

Implementing Security Technical Implementation Guide (STIG) controls remains a demanding operating-system administration task. Each natural-language security requirement must be translated into a precise and persistent configuration change. These changes span Secure Shell (SSH) configuration, authentication policy, audit rules, and kernel parameters. Existing benchmarks do not measure whether a model can perform this task on a live system. Code benchmarks evaluate isolated programs without persistent system state, while agent benchmarks largely focus on applications rather than operating-system configuration. In this paper, we introduce \textbf{HardEval}, an execution-verified benchmark for evaluating language-model-assisted security hardening. HardEval is built from Red Hat Enterprise Linux 8 STIG controls that fail on an unhardened baseline system. Expert-written reference remediations and Open Vulnerability and Assessment Language (OVAL) checks serve as hidden ground truth for grading; they are never shown to the model, and each task states only the required security outcome while withholding the implementation. The model produces a Bash script, which is executed on a live virtual machine and graded from fail to pass using the corresponding OpenSCAP/OVAL compliance check. Evaluation of fourteen frontier and open-weight configurations reveals large and uneven capability differences across security domains. A recurring failure mode is plausible security code that does not satisfy the exact compliance condition. For four open-weight configurations, multi-run Wilson confidence intervals preserve the observed ordering. HardEval is publicly available.

\end{abstract}


\section{Introduction}
\label{sec:intro}

\begin{figure*}[t]
\centering
\includegraphics[width=0.90\textwidth]{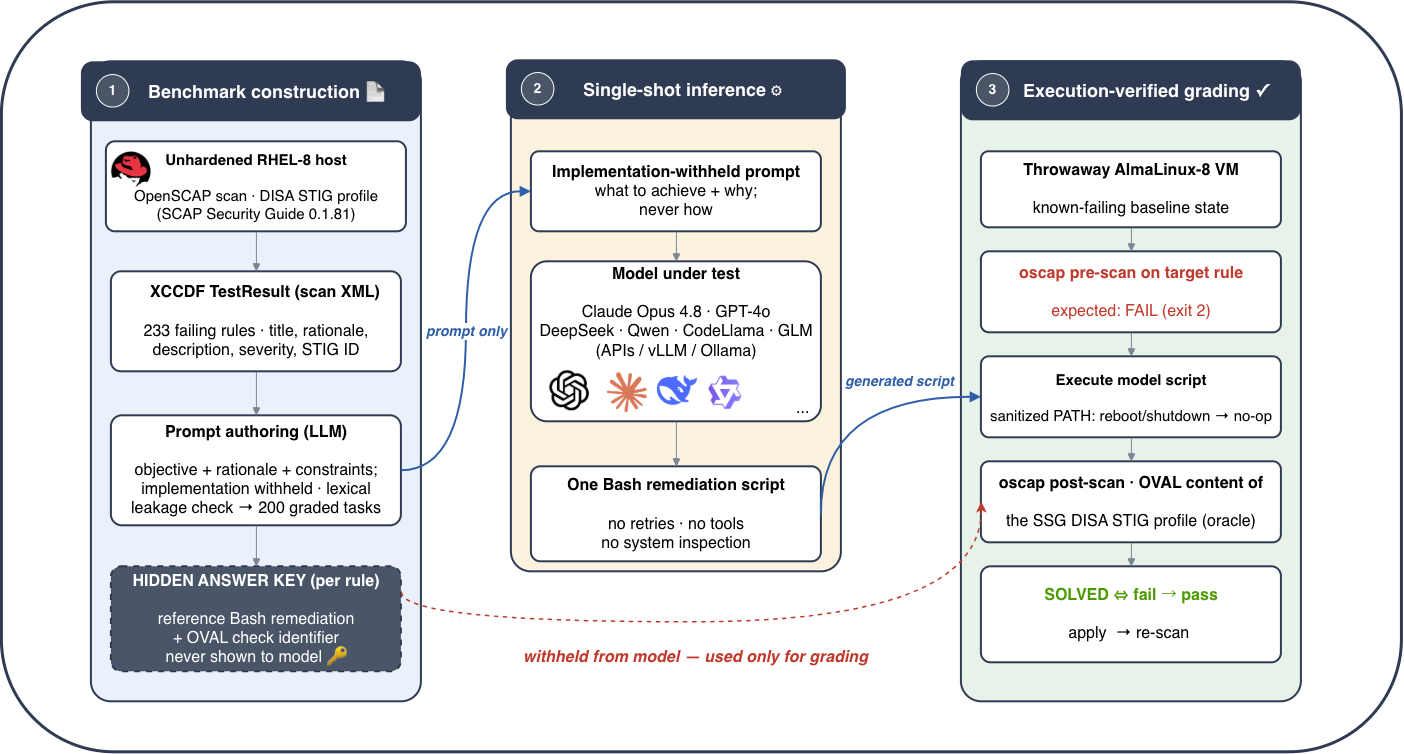}
\caption{HardEval system overview. Stage~1 turns each rule failing an OpenSCAP scan of an unhardened RHEL-8 host into an implementation-withheld task plus a hidden answer key (reference remediation and OVAL check identifier). Stage~2 gives the model only the prompt, and Stage~3 executes the returned Bash script on a throwaway VM and re-runs the same OVAL check: a task is solved iff the rule flips from \texttt{fail} to \texttt{pass}.}
\label{fig:system-overview}
\end{figure*}

Operating-system misconfigurations remain a persistent source of cyber risk. In Department of Defense (DoD) and regulated enterprise environments, this risk is managed through hardening frameworks such as the Defense Information Systems Agency (DISA) Security Technical Implementation Guides (STIGs). In practice, STIG compliance requires administrators to translate hundreds of natural-language security requirements into precise, durable configuration changes spanning kernel parameters, authentication policy, audit rules, SSH, services, package state, and reboot-persistent system state. Tools such as OpenSCAP, ComplianceAsCode, Ansible Lockdown, and commercial products can automate scanning and apply pre-authored remediations, but they remain constrained by static rule content and manually maintained corrective actions. When no suitable fix exists, when configurations deviate from expected templates, or when policy guidance evolves, the burden falls back to human analysts and administrators.

Large language models (LLMs) suggest a more adaptive alternative to static remediation playbooks: instead of selecting from prewritten remediations, a model may generate remediation code directly from a natural-language security requirement. However, end-to-end remediation systems can conflate model capability with retrieval context, post-processing, and orchestration. HardEval instead isolates the underlying code-generation capability: whether a model can produce security-hardening code that makes a live operating system compliant in the exact way an external scanner verifies. An isolated measurement also serves model selection: before investing in fine-tuning or retrieval pipelines, practitioners, especially those who must self-host, need a way to identify which base model can already bring a system to scanner-verified compliance.

Existing evaluation suites are poorly matched to this question: execution-based code benchmarks score isolated snippets with no persistent system state, repository- and agent-grounded benchmarks target application code rather than OS configuration, and security-oriented evaluations leave OS hardening unmeasured; even the most closely related system, SHIELDS~\cite{Hamara2026SHIELDS}, evaluates an end-to-end remediation loop rather than isolating the code-generation capability it requires (Section~\ref{sec:related}).

This paper addresses that gap with \textbf{HardEval}, a functionally verified benchmark for LLM-generated Linux security-hardening code on the RHEL~8 family. HardEval is built from failed, automatable DISA STIG controls and expert-maintained ComplianceAsCode security content, including authoritative OVAL checks and reference remediation logic. Each task is presented to the model as an implementation-withheld natural-language security requirement that states the hardening objective and rationale while withholding the exact mechanism. The model must generate a Bash remediation script, which is then executed on a fresh RHEL-compatible virtual machine and scored by re-running the corresponding OpenSCAP/OVAL rule. A task is counted as solved only if the rule transitions from \texttt{fail} to \texttt{pass}. This target, compliance-oracle satisfaction, is deliberately narrower than security in the broad sense (Section~\ref{sec:hardeval-protocol});

Fig.~\ref{fig:system-overview} summarizes the end-to-end design.

On the fixed cross-model manifest, the strongest configuration, Claude Opus~4.7, achieves 111/137 (81.0\%) on the combined server-safe bucket, with Claude Opus~4.8 at 106/137 (77.4\%).  Failures are rarely shell-syntax errors; the recurring patterns are scanner-specific value or mechanism mismatches. This result points to a key distinction between plausible security code and verifiable compliance: a script may improve security in practice while still failing the exact condition the scanner checks.

The contributions of this paper are as follows:
\begin{itemize}
    \item We formalize a new evaluation problem, whether LLMs can generate remediation code that brings live operating systems into compliance under authoritative OpenSCAP/OVAL checks, and introduce \textbf{HardEval}, a reusable execution-based benchmark built from DISA STIG failures and ComplianceAsCode content, with implementation-withheld prompts and fail-to-pass grading on throwaway virtual machines.
    \item We provide a baseline study spanning fourteen model configurations, from frontier closed models to 4-bit quantized open weights, showing that frontier models solve many hardening tasks but fail on scanner-sensitive details invisible to text-based evaluation.
    \item We validate the benchmark's robustness with a cross-author study (GPT-4o re-authoring the prompts) and a zero-code port to Ubuntu~24.04, showing that the model ordering is stable across prompt authors and that the construction pipeline is not tied to a single operating system.
\end{itemize}

\section{Related Work}
\label{sec:related}

Classical code-generation benchmarks such as HumanEval~\cite{Chen2021Eval} and MBPP~\cite{Austin2021} evaluate Python function synthesis from docstrings via unit-test execution, and APPS~\cite{Hendrycks2021APPS} extends this to 10{,}000 coding-interview problems. MultiPL-E~\cite{Cassano2022MultiPLE} adapts these suites to 18 programming languages, BigCodeBench~\cite{zhuo2025bigcodebench} targets complex tasks requiring diverse library calls, and LiveCodeBench~\cite{Jain2024LiveCodeBench} draws on recent contest problems for contamination-resistant evaluation. All of these are execution-based, but they evaluate isolated snippets with no persistent system state and no security or OS-configuration concerns.

A second line of work grounds evaluation in realistic environments. SWE-bench~\cite{Jimenez2024SWEbench} requires models to edit real repositories so that failing tests pass, a protocol it calls fail-to-pass. WebArena~\cite{Zhou2023WebArena}, OSWorld~\cite{Xie2024OSWorld}, and AgentBench~\cite{liu2024agentbench} evaluate agents on end-to-end web, desktop, and simulated-environment tasks. None of these benchmarks test the generation of OS-hardening shell scripts against domain-specific compliance checks, and many permit agent-tool loops that could trivially reveal the fix. HardEval instead uses implementation-withheld prompts, so the exact remediation is never exposed to the model.

Security-focused evaluations emphasize either insecure-code avoidance or offensive capability. CyberSecEval~3~\cite{Wan2024CyberSecEval3} measures cyber risks and capabilities such as exploit generation and phishing. LLMSecEval~\cite{Tony2023LLMSecEval} provides 150 prompts describing CWE-vulnerable code. HardSecBench~\cite{Chen2026HardSecBench} evaluates security-aware hardware and firmware code against executable oracles. Closest to our work, SHIELDS~\cite{Hamara2026SHIELDS} remediates STIG findings through an iterative multi-agent loop driven by scanner feedback, resolving up to 73\% of findings. %
SHIELDS evaluates an end-to-end agentic system, whereas HardEval is a reusable benchmark of single-shot remediation-code generation (no agent orchestration, no scanner feedback) under fail-to-pass OpenSCAP scoring. Table~\ref{tab:benchmarks} situates HardEval against representative benchmarks along the dimensions that matter for OS hardening. To our knowledge, HardEval is the first benchmark whose target artifact is OS-hardening Bash graded by execution and compliance-scanner verification on live virtual machines.

\begin{table*}[t]
\centering
\caption{Comparison of Representative Benchmarks. ``Exec'' = Whether Grading Executes the Model's Output; ``Env'' = the Environment That Output Runs In; ``Tools'' = External Tool/Agent Access During the Task.}
\label{tab:benchmarks}
\footnotesize
\setlength{\tabcolsep}{5pt}
\begin{adjustbox}{max width=\textwidth}
\begin{tabular}{@{}llllll@{}}
\toprule
\textbf{Benchmark} & \textbf{Domain} & \textbf{Exec} & \textbf{Env} & \textbf{Tools} &  \textbf{Target artifact} \\
\midrule
HumanEval \cite{Chen2021Eval}              & Python functions           & Yes (tests)         & None          & None                          & Python code \\
MBPP \cite{Austin2021}                     & Python script tasks        & Yes (tests)         & None          & None                          & Python code \\
BigCodeBench \cite{zhuo2025bigcodebench}   & Complex code (multi-call)  & Yes (tests)         & None          & None                          & Python code \\
LiveCodeBench \cite{Jain2024LiveCodeBench} & Contest problems           & Yes (tests)         & None          & None                          & Python code \\
\midrule
SWE-bench \cite{Jimenez2024SWEbench}       & GitHub issue repair        & Yes (fail$\to$pass) & Repo (Docker) & Code editing                  & Repository patch \\
WebArena \cite{Zhou2023WebArena}           & Web app tasks              & Yes (sim.)          & Web simulator & Extensive web tools           & Action sequence \\
OSWorld \cite{Xie2024OSWorld}              & Desktop tasks              & Yes (sim.)          & Real OS       & GUI/OS tools                  & Multi-step actions \\
\midrule
LLMSecEval \cite{Tony2023LLMSecEval}       & Insecure-code prompts      & No                  & None          & None                          & Code snippets \\
HardSecBench \cite{Chen2026HardSecBench}   & Hardware/firmware code     & Yes (tests)         & Simulated     & None                          & RTL/C code \\
\midrule
\textbf{HardEval (ours)}                   & OS hardening               & Yes (OpenSCAP)      & Real VMs      & None                          & Bash scripts \\
\bottomrule
\end{tabular}
\end{adjustbox}
\end{table*}

\section{The HardEval Benchmark}
\label{sec:hardeval-design}

In this section, we describe the source material and dataset construction, the implementation-withheld prompt authoring process, the execution-based evaluation protocol, and the models and experimental setup. Together, they form the pipeline that isolates hardening capability from retrieval, orchestration, and leakage confounds.

\subsection{Source Material and Dataset Construction}
\label{sec:hardeval-source}

HardEval is bootstrapped from a single source of truth. This source is an OpenSCAP scan of an unhardened RHEL~8 host executed under the DISA STIG profile using SCAP Security Guide~0.1.81 from the ComplianceAsCode/content project. This scan produces a fully rendered XCCDF \texttt{TestResult} document that provides, for each failing rule, three things: (i)~the rule title, description, and rationale authored by Red Hat security engineers; (ii)~the reference Bash remediation script, which constitutes the hidden answer key; and (iii)~the OVAL check identifier, which is the grading oracle. Rule identifiers are cross-mapped to official DISA STIG IDs, with NIST~800-53, SRG, CCE, and CCI references retained as metadata.

Dataset construction proceeds in two decoupled stages. An \emph{extraction} stage parses the scan XML offline into reference data and grading metadata, and a \emph{generation} stage authors the implementation-withheld task prompts (Section~\ref{sec:hardeval-prompts}). From the 233 rules that fail the initial scan, we select the 215 that carry a reference Bash remediation. The 200 of these that are automatable constitute the \emph{graded tasks} (Section~\ref{sec:hardeval}). A fixed cross-model \emph{manifest} of 151 verified-applicable rules (137 excluding SSH) is graded identically for every model (Section~\ref{sec:hardeval}). This manifest omits reboot-persistent kernel rules, hazardous access-severing rules, and GUI/hardware-dependent controls that cannot be evaluated to completion on a single live VM. Each item is a tuple $(p_i,\, r_i,\, o_i,\, \textit{stig\_id},\, \textit{sev},\, \textit{reboot})$, consisting of the prompt, the hidden reference remediation, the OVAL check identifier, and the STIG ID, severity, and reboot-persistence flags. The prompt $p_i$ is the only element exposed to the model. $r_i$ and $o_i$ are withheld at inference time and used only during grading.

\subsection{Implementation-Withheld Prompt Authoring}
\label{sec:hardeval-prompts}

Each prompt $p_i$ is authored by a generative model (Claude Opus~4.8, queried through the Batch API) conditioned on the rule's title, severity, rationale, and description text, but never on the reference remediation script. ComplianceAsCode descriptions frequently embed implementation specifics, for example ``Set \texttt{ClientAliveInterval} to 600 in \texttt{/etc/ssh/sshd\_config}.'' Because of this, the authoring model operates under a structured system prompt, calibrated with few-shot exemplars and constrained to a JSON output schema, that enforces four requirements:

\begin{enumerate}
  \item \textbf{State the objective and rationale} in plain English. The required end state must be precise enough that the OVAL check can grade it against one oracle-accepted outcome, and the threat or rationale must ground the objective.
  \item \textbf{Omit the implementation mechanism.} Exact paths, directive and parameter names, \texttt{sysctl} keys, shell commands, and literal configuration lines are abstracted away; naming the affected component conceptually (``the SSH server'') is permitted, naming the knob, path, or command is not. Required target \emph{values or thresholds} (``a minimum length of 15 characters'') are preserved.
  \item \textbf{Preserve correctness constraints} that are mechanism-independent, including idempotency, persistence across reboots, and scope, and append a fixed output-format instruction requiring a single self-contained Bash script in one fenced code block.
  \item \textbf{Run a programmatic lexical-leakage check.} A whole-word, case-insensitive scan compares each prompt against mechanism tokens extracted from the hidden reference script (paths, \texttt{sysctl}-style keys, directive names); of 215 prompts, 2 were flagged and both confirmed benign.
\end{enumerate}

The generator system prompt, exemplars, and output-format instruction are in the repository.

Fig.~\ref{fig:prompt-example} shows the prompt for \texttt{accounts\_password\_pam\_minlen} (RHEL-08-020230): the required length is disclosed while the target file, directive, and enforcement mechanism are withheld. A model that knows Linux hardening can solve this; one that merely transcribes reference scripts cannot, because the script is never shown.

\begin{figure}[t]
\centering
\begin{promptpanel}{HardEval Task Prompt \hfill RHEL-08-020230 \,\(\cdot\)\, CAT II (medium)}
\footnotesize
\textbf{\sffamily Requirement.}
RHEL~8 must enforce a minimum password length of 15 characters. Short passwords are trivially guessable and provide inadequate protection against brute-force attacks. Enforcing a minimum length reduces the probability that user-chosen passwords will be weak.

\smallskip
\textbf{\sffamily Objective.}
Configure the operating system so that all new passwords set or changed through the standard system authentication infrastructure satisfy a minimum length of 15 characters. This restriction must apply to all interactive users and must persist across reboots.

\end{promptpanel}
\caption{Example implementation-withheld HardEval prompt (RHEL-08-020230): \emph{what} must be achieved is stated; \emph{how} (target file, directive, PAM module) is withheld. The hidden reference writes \texttt{minlen = 15} to \texttt{/etc/security/pwquality.conf}, which the OVAL oracle checks at grading time. The constraints and output-format fields are omitted here (full prompt text in the repository).}
\label{fig:prompt-example}
\end{figure}

\subsection{Evaluation Protocol}
\label{sec:hardeval-protocol}

HardEval employs a strict single-shot, execution-based grading protocol:

\begin{enumerate}
    \item The model receives only the implementation-withheld prompt and emits one Bash script (no retries, no tool access, no live system inspection).
    \item Each configuration is graded on its own throwaway RHEL-8-compatible virtual machine (AlmaLinux~8), restored to the identical known-failing baseline, with no other model's code ever touching it. Scripts are applied sequentially, each preceded by a diagnostic pre-scan of its target rule.
    \item The model's script is executed against a sanitized \texttt{PATH} that neutralizes \texttt{reboot}, \texttt{shutdown}, \texttt{poweroff}, and \texttt{systemctl reboot} to prevent the model from rebooting the test host mid-run.
    \item OpenSCAP is re-invoked on the target rule, and the verdict is parsed from the reported rule result, not the process exit code: only the literal \texttt{pass} is credited, never \texttt{notapplicable} or another non-failing state. After the final script, a full re-scan of every manifest rule verifies the end state; in all retained RHEL runs it agreed with the rule-level outcomes, with no credited pass reversed and no graded rule broken.
\end{enumerate}

Because of this, there is no reward for plausible-looking but non-compliant code and no string-similarity confound. The graded quantity is the end-state compliance of the machine hardened by a configuration's scripts: scripts persist on the VM across a run, so HardEval measures cumulative batch hardening; the aggregate is exact, while credit for an individual rule is not always attributable to the script written for its task. The harness assumes non-adversarial models (the scoring logic and hidden references reside outside the evaluated machine, and a scan of all 6{,}494 generated scripts found no reference to the scanner, the SCAP datastream, or any grading artifact); a script that instead disables an entire scoring subsystem surfaces as an aggregate anomaly and its run is excluded as invalid (Section~\ref{sec:cross-os}).

\paragraph{Scope: compliance-oracle satisfaction, not security in the broad sense}
HardEval's unit of measurement is deliberately narrow: it measures whether LLM-generated remediation \emph{satisfies the DISA STIG compliance oracle}, not whether the system is ``more secure'' in an unrestricted sense. A script may improve security through an alternative mechanism yet still fail the probe (Section~\ref{sec:predictions}), and a target-rule pass by itself does not certify the absence of collateral effects elsewhere on the host. We adopt this framing because it keeps grading objective and reproducible: there is exactly one adjudicator, and it is external to both the model and the authors, and because it is operationally faithful: in accredited DoD and enterprise environments, compliance is adjudicated by precisely these scanner checks, so oracle satisfaction is the deployable quantity.

\subsection{Models and Experimental Setup}
\label{sec:hardeval-models}

We evaluate fourteen model configurations. Seven are frontier closed models, queried through their official APIs: \textbf{Claude Opus~4.8}, \textbf{Claude Opus~4.7}, \textbf{Claude Opus~4.5}, \textbf{Claude Sonnet~5}, \textbf{Claude Sonnet~4.5}, \textbf{Claude Haiku~4.5}, and \textbf{GPT-4o}. Seven are open-weight, code-specialized configurations: \textbf{DeepSeek-Coder-33B-Instruct}, \textbf{CodeLlama-34B-Instruct}, \textbf{CodeLlama-7B-Instruct}, \textbf{Qwen2.5-Coder-14B-Instruct}, \textbf{Qwen2.5-Coder-7B-Instruct}, and \textbf{GLM-4-9B} in both FP16 and 4-bit form, the pair included to measure the cost of aggressive quantization. The Qwen models were served via vLLM on a single NVIDIA A100 GPU; the remaining open-weight models via Ollama at FP16, with the 4-bit GLM-4-9B using Ollama's default \texttt{Q4\_K\_M} quantization. Every configuration received the same prompts and decoded greedily at temperature~0, and every script was scored, separately from inference, by the same model-agnostic grader on an identical AlmaLinux~8 host (OpenSCAP 1.3.14, SCAP Security Guide 0.1.81, \texttt{stig} profile): it ingests a \texttt{predictions\_<model>.jsonl} file and extracts the script from a \texttt{```bash} fence, any fenced block, or raw text as fallback, so no model is penalized for output formatting.

The complete construction pipeline described in this section is implemented in \texttt{benchmark/build\_dataset.py} in the ``code and data supplimentary" folder.

\section{Evaluation and Results}
\label{sec:evaluation}

We report cross-model HardEval results (Section~\ref{sec:hardeval}), run-to-run variance (Section~\ref{sec:variance}), cross-author prompt validation and cross-OS generalizability (Sections~\ref{sec:cross-author}--\ref{sec:cross-os}), and failure analysis (Section~\ref{sec:predictions}).

\subsection{Cross-Model HardEval Results}
\label{sec:hardeval}

Table~\ref{tab:hardeval-main} reports HardEval results across all fourteen model configurations, broken down by the principal evaluation buckets: server configuration plus kernel hardening, audit rules, the combined server-safe aggregate (the headline metric), SSH configuration, and all verified-applicable rules.  Invalid, empty, or unscored outputs are counted as failures, and every configuration shares identical fixed denominators (137 server-safe, 151 verified-applicable). %
The results reveal four findings.

\begin{table*}[t]
\centering
\caption{HardEval Main Results: Manifest Rules Brought from Fail to Pass Under OpenSCAP/OVAL Grading, with Fixed 137/151 Denominators for All Configurations (Section~\ref{sec:limitations}). ``Combined server-safe'' pools the server-configuration and audit-rule buckets; ``all verified applicable'' adds SSH; GUI/hardware rules are non-applicable on the headless host. Invalid or unscored outputs count as failures. No collateral breakage was observed among the 151 manifest rules (final full re-scan; Section~\ref{sec:hardeval-protocol}); controls outside the manifest and general system functionality were not evaluated.}
\label{tab:hardeval-main}
\footnotesize
\setlength{\tabcolsep}{7pt}
\begin{adjustbox}{max width=\linewidth}
\begin{tabular}{@{}lccccc@{}}
\toprule
\textbf{Model} & \textbf{Server config + kernel} & \textbf{Audit rules} & \textbf{Combined server-safe} & \textbf{SSH config} & \textbf{All verified applicable} \\
\midrule
Claude Opus 4.7            & 70/79 (\textbf{88.6\%})   & 41/58 (70.7\%)          & \textbf{111/137 (81.0\%)} & 11/14 (\textbf{78.6\%}) & \textbf{122/151 (80.8\%)} \\
Claude Opus 4.8            & 65/79 (82.3\%)          & 41/58 (70.7\%)          & 106/137 (77.4\%)          & 10/14 (71.4\%)          & 116/151 (76.8\%) \\
Claude Sonnet 4.5          & 58/79 (73.4\%)            & 47/58 (\textbf{81.0\%}) & 105/137 (76.6\%)          & 9/14 (64.3\%)           & 114/151 (75.5\%) \\
Claude Opus 4.5            & 66/79 (83.5\%)            & 38/58 (65.5\%)          & 104/137 (75.9\%)          & 6/14 (42.9\%)           & 110/151 (72.8\%) \\
GPT-4o                     & 54/79 (68.4\%)            & 36/58 (62.1\%)          & 90/137 (65.7\%)           & 10/14 (71.4\%)          & 100/151 (66.2\%) \\
Claude Sonnet 5            & 48/79 (60.8\%)            & 34/58 (58.6\%)          & 82/137 (59.9\%)           & 7/14 (50.0\%)           & 89/151 (58.9\%) \\
Claude Haiku 4.5           & 49/79 (62.0\%)            & 21/58 (36.2\%)          & 70/137 (51.1\%)           & 6/14 (42.9\%)           & 76/151 (50.3\%) \\
DeepSeek-Coder-33B (FP16)  & 35/79 (44.3\%)            & 14/58 (24.1\%)          & 49/137 (35.8\%)           & 6/14 (42.9\%)           & 55/151 (36.4\%) \\
Qwen2.5-Coder-14B          & 32/79 (40.5\%)            & 14/58 (24.1\%)          & 46/137 (33.6\%)           & 7/14 (50.0\%)           & 53/151 (35.1\%) \\
CodeLlama-34B (FP16)       & 20/79 (25.3\%)            & 9/58 (15.5\%)           & 29/137 (21.2\%)           & 3/14 (21.4\%)           & 32/151 (21.2\%) \\
Qwen2.5-Coder-7B           & 19/79 (24.1\%)            & 1/58 (1.7\%)            & 20/137 (14.6\%)           & 4/14 (28.6\%)           & 24/151 (15.9\%) \\
CodeLlama-7B (FP16)        & 16/79 (20.3\%)            & 4/58 (6.9\%)            & 20/137 (14.6\%)           & 5/14 (35.7\%)           & 25/151 (16.6\%) \\
GLM-4-9B (FP16)            & 16/79 (20.3\%)            & 2/58 (3.4\%)            & 18/137 (13.1\%)           & 3/14 (21.4\%)           & 21/151 (13.9\%) \\
GLM-4-9B (4-bit)           & 17/79 (21.5\%)            & 0/58 (0.0\%)            & 17/137 (12.4\%)           & 2/14 (14.3\%)           & 19/151 (12.6\%) \\
\bottomrule
\end{tabular}
\end{adjustbox}
\end{table*}

\paragraph{Frontier-scale models dominate, but scale is only a coarse predictor}
On the fixed cross-model manifest, Claude Opus~4.7 leads with 81.0\% verified success on the combined server-safe bucket, with Claude Opus~4.8 at 77.4\%, a single-run gap we do not read as an ordering between the two (Section~\ref{sec:variance}). The seven frontier models, whose undisclosed parameter counts are generally understood to be far larger than the 7B--34B open-weight models evaluated here, occupy the top seven positions (81.0\% down to 51.1\% for Haiku~4.5, with GPT-4o at 65.7\% among them). The strongest open-weight configuration, DeepSeek-Coder-33B at 35.8\%, trails the weakest frontier model by more than fifteen points, and code specialization alone does not close the gap. Scale separates the tiers but does not order models within them, as the next finding shows.

\paragraph{Within the evaluated set, parameter count is a weak predictor of hardening ability}
CodeLlama-34B (21.2\%) trails the far smaller Qwen2.5-Coder-14B (33.6\%), and DeepSeek-Coder-33B (35.8\%) exceeds it only narrowly, despite both carrying more than twice the parameters. The Qwen-14B/CodeLlama-34B gap survives repeated sampling, since their pooled multi-run confidence intervals (Section~\ref{sec:variance}) do not overlap. Training recency and data composition appear to dominate raw scale here, though model family, instruction tuning, and decoding remain uncontrolled confounders across families. Release recency within a family is likewise not a reliable predictor: Claude Opus~4.8 does not outscore Opus~4.7 on a single run, and Sonnet~5 trails the older Sonnet~4.5 by more than sixteen points.

\paragraph{Capability is domain-structured: audit rules discriminate most sharply}
Performance on \texttt{auditd} rule construction separates models more sharply than any other bucket: Qwen~7B achieves 1.7\% on audit rules versus 24.1\% on server configuration, GLM-4-9B manages 3.4\% in FP16 and 0\% at 4-bit, and even GPT-4o lags (62.1\% vs.\ 70.7\% for Claude Opus~4.8 and 81.0\% for Claude Sonnet~4.5). Nor does capability transfer uniformly across domains: GPT-4o's SSH score (71.4\%) matches Claude Opus~4.8 (though SSH carries a measurement-channel confound; Sections~\ref{sec:predictions} and~\ref{sec:limitations}), Qwen-14B's SSH success (50.0\%) doubles its own audit-rule rate, and Sonnet~4.5 posts the highest audit-rule score of any model (81.0\%) while ranking only third on the combined metric. These domain-level inversions would be invisible to an aggregate compliance score.

\paragraph{No detectable quantization cost for GLM-4-9B at this sample size}
Under Ollama's \texttt{Q4\_K\_M} 4-bit quantization, GLM-4-9B's aggregate accuracy barely moves, from 13.1\% to 12.4\% single-run, and the multi-run study confirms it. Pooled estimates of 14.1\% (FP16) and 12.2\% (4-bit) carry overlapping confidence intervals (Section~\ref{sec:variance}). The single-run audit-rule score is very low in both cases, 3.4\% for FP16 and 0.0\% for 4-bit quantization. The pooled multi-run audit estimates, 1.3\% and 0.4\%, overlap, indicating a family-level weakness on \texttt{auditd} grammar rather than a quantization-specific collapse.

\subsection{Run-to-Run Variance}
\label{sec:variance}

\begin{figure}[t]
\centering
\includegraphics[width=\columnwidth]{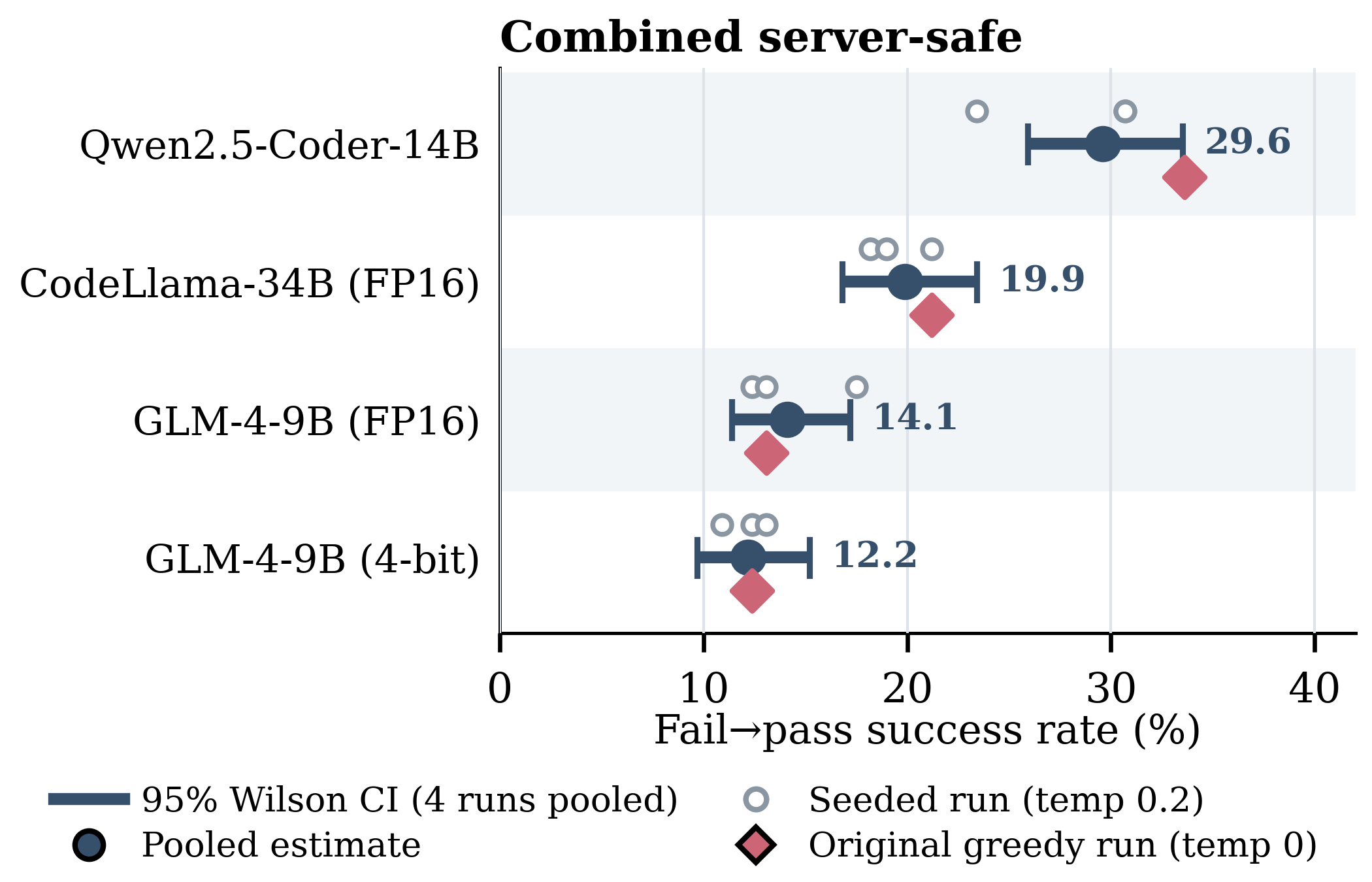}
\caption{Run-to-run variance for four open-weight configurations: 95\% Wilson intervals pooling one greedy (temperature-0) run with three seeded runs at temperature 0.2, on the combined server-safe metric (seeded runs: open circles; greedy run: red diamonds). Qwen-14B and CodeLlama-34B do not overlap; the two GLM-4-9B precisions overlap.}
\label{fig:ci-forest}
\end{figure}

A single greedy generation is one sample. For four open-weight configurations we therefore pooled the temperature-0 run with three independent seeded runs (temperature 0.2) and computed 95\% Wilson score intervals over the combined server-safe metric, treating each (rule, run) pair as a Bernoulli trial ($4 \times 137 = 548$ trials per model; Fig.~\ref{fig:ci-forest}). We read the intervals descriptively, as a sensitivity analysis rather than formal inference: the pool mixes two decoding settings, and trials on the same rule are correlated across runs, so the intervals are, if anything, slightly narrow. Every pooled estimate preserves the single-run ranking, and greedy runs read mildly optimistic: Qwen-14B's greedy run (33.6\%) lies just outside its pooled interval (29.6\%, 95\% CI 25.9--33.5\%), while the other three greedy runs fall inside theirs. The Qwen-14B and CodeLlama-34B intervals remain separated, consistent with the parameter-count finding, whereas the two GLM-4-9B precisions overlap on both the combined metric and audit rules (1.3\% vs.\ 0.4\% pooled), indicating no detectable quantization penalty at this sample size.

\subsection{Cross-Author Prompt Validation}
\label{sec:cross-author}

Because the main prompt set is authored by Claude Opus~4.8 (Section~\ref{sec:hardeval-prompts}), a natural concern is that its prompts might advantage Claude as a solver. To test whether the authoring process and its implementation-withheld result are specific to Claude or hold for an independently authored set, GPT-4o authored a parallel prompt set from the \emph{same} raw STIG metadata (title, severity, rationale, description), under the identical instructions and lexical-leakage checker (\texttt{benchmark/generate\_prompts\_gpt.py}). GPT-4o's initial pass flagged 13 of 215 rows for possible leakage, against 0 genuine leaks for Claude; 6 were confirmed real mechanism leaks (exact file paths or configuration module names) and regenerated. Both prompt sets were then run and scored identically, using the same harness, the same AlmaLinux~8 host, and the same OVAL grading. Table~\ref{tab:cross-author} reports both solvers on the GPT-4o-authored set.

\begin{table}[t]
\centering
\caption{Cross-Author Validation: Claude Opus~4.8 and GPT-4o Solving the GPT-4o-Authored Prompt Set. The prompts cover the same underlying rules under the same fixed 137-rule (SSH-excluded) denominators; for reference, on the original Claude-authored prompts the same solvers score 77.4\% and 65.7\% combined.}
\label{tab:cross-author}
\footnotesize
\setlength{\tabcolsep}{4pt}
\begin{adjustbox}{max width=\linewidth}
\begin{tabular}{@{}lccc@{}}
\toprule
\textbf{Model} & \textbf{Server+kernel} & \textbf{Audit} & \textbf{Combined} \\
\midrule
Claude Opus 4.8 & 60/79 (75.9\%) & 41/58 (70.7\%) & 101/137 (73.7\%) \\
GPT-4o          & 43/79 (54.4\%) & 30/58 (51.7\%) & 73/137 (53.3\%)  \\
\bottomrule
\end{tabular}
\end{adjustbox}
\end{table}

The result is the opposite of a self-authoring advantage: both solvers score \emph{lower} on the GPT-4o-authored prompts (Claude Opus~4.8 falls from 77.4\% to 73.7\%, GPT-4o from 65.7\% to 53.3\%), and Claude remains ahead on either author's set. Claude-authored prompts yield higher solve rates from \emph{both} solvers, indicating that prompt clarity and specificity matter more than any author--solver relationship. We read this as evidence that the model ordering is author-robust on these two prompt sets, not that absolute scores are prompt-invariant, addressing the authoring-bias concern in Section~\ref{sec:limitations}.

\subsection{Cross-OS Portability}
\label{sec:cross-os}

HardEval ports from RHEL~8 to Ubuntu~24.04 with no code changes: only the OVAL datastream (\texttt{ssg-rhel8-ds.xml} $\to$ \texttt{ssg-ubuntu2404-ds.xml}) and the OS name in the prompt text change. The harness uses the OS-specific STIG content installed on each VM rather than embedding RHEL-specific remediation logic. Models inferred the appropriate tooling from the OS name alone (\texttt{apt}/\texttt{dpkg} vs.\ \texttt{yum}/\texttt{rpm}).

\begin{table}[t]
\centering
\caption{Cross-OS Portability: HardEval Ported to Ubuntu~24.04 (88 Rules Shared Between the RHEL~8 and Ubuntu STIGs). Only the three clean end-to-end runs are shown; two contaminated runs are described in the text.}
\label{tab:cross-os}
\footnotesize
\setlength{\tabcolsep}{4pt}
\begin{adjustbox}{max width=\linewidth}
\begin{tabular}{@{}lcccc@{}}
\toprule
\textbf{Model} & \textbf{Server+kernel} & \textbf{Audit} & \textbf{SSH} & \textbf{Total} \\
\midrule
Claude Opus 4.8 & 14/35 (40.0\%) & 42/47 (89.4\%) & 4/6 (66.7\%) & 60/88 (\textbf{68.2\%}) \\
GPT-4o          & 15/35 (42.9\%) & 36/47 (76.6\%) & 4/6 (66.7\%) & 55/88 (62.5\%) \\
Claude Sonnet 5 & 16/35 (45.7\%) & 35/47 (74.5\%) & 2/6 (33.3\%) & 53/88 (60.2\%) \\
\bottomrule
\end{tabular}
\end{adjustbox}
\end{table}

Table~\ref{tab:cross-os} reports the three models with valid end-to-end runs, each evaluated on a separate clean Ubuntu~24.04 VM. Two exploratory runs were invalidated by a scoring cascade in which one script broke the audit subsystem through \texttt{sssd\_enable\_smartcards}, causing all 47 subsequently evaluated \texttt{audit\_rules\_*} checks to fail. The triggering script is itself a genuine model failure on its own rule, but the 47 poisoned verdicts are measurement artifacts, so the runs are excluded as invalid rather than scored; their per-rule results are retained in the repository.

The ported pipeline produces non-degenerate and well-separated pass rates across the three valid runs. The 88 rules are every scoreable concept the two STIGs share, not a hand-picked subset; the rest depend on mechanisms absent from Ubuntu (SELinux, \texttt{authselect}) or lack an Ubuntu counterpart.

These results establish the portability of HardEval's construction and execution-based grading pipeline, but not the invariance of task difficulty across operating systems. The domain profile changes on Ubuntu: audit rules remain highly solvable for the strongest models, whereas server and kernel configuration becomes harder. HardEval's domain labels are therefore operational categories, not invariant difficulty classes.

\subsection{Failure Analysis}
\label{sec:predictions}

HardEval's rule-level verdicts make failures directly inspectable: a rule that ends the run failing was fixed by no script in the batch. On the Fig.~\ref{fig:prompt-example} task, for example, Claude Opus~4.8, GPT-4o, and Qwen2.5-Coder-14B all locate \texttt{/etc/security/pwquality.conf} and pass, while Qwen2.5-Coder-7B writes \texttt{PASS\_MIN\_LEN} to \texttt{/etc/login.defs}, a valid-looking legacy control the OVAL probe does not inspect, and fails.

Across all models, we observed three recurring failure patterns. Fig.~\ref{fig:failure-modes} summarizes these. Their shared structure is precisely the gap --- between \emph{plausible hardening code} and \emph{verifiably compliant configuration} --- that HardEval is designed to measure (see the scope discussion in Section~\ref{sec:hardeval-protocol}), and that no text-similarity or LLM-judge evaluation can detect.

\begin{figure}[t]
\centering
\begin{promptpanel}{Recurring Failure Modes \hfill examples: Claude Opus~4.8}
\footnotesize
\textbf{\sffamily Value mismatch.}
The script sets a sensible, security-improving value (e.g., 5{,}000 SHA-512 rounds); the OVAL probe requires exactly 100{,}000 rounds and flags anything lower as \texttt{fail}.

\textbf{\sffamily Mechanism mismatch.}
The correct security property is configured via a valid-but-wrong mechanism (e.g., PAM \texttt{faildelay} for login delay); OVAL checks \texttt{FAIL\_DELAY} in \texttt{/etc/login.defs} and never inspects the PAM change.

\textbf{\sffamily Audit key mismatch.}
A syntactically correct \texttt{auditd} rule targets the right syscall; OVAL probes for an exact \texttt{-k} key string (e.g., \texttt{privileged-chacl}), so any other key is scored \texttt{fail}.
\end{promptpanel}
\caption{Recurring failure modes on HardEval, observed across all fourteen configurations: in each, the script arguably improves security but misses the exact file, directive, or value the OVAL probe checks. }
\label{fig:failure-modes}
\end{figure}
Per-domain failure rates, derivable from Table~\ref{tab:hardeval-main} (per-rule results for all fourteen configurations are in the repository), show three patterns. First, audit-rule construction is the sharpest discriminator (Section~\ref{sec:hardeval}): \texttt{auditd} rule syntax is a scarce skill concentrated in frontier models. Second, SSH results require separate interpretation because aggressive SSH hardening can sever the grading channel itself (Section~\ref{sec:limitations}); the inversion suggested by the earlier extended run did not reproduce in the manifest re-run, where GPT-4o and Claude Opus~4.8 both score 10/14. Third, package and service rules show the shallowest gradient, likely because common \texttt{dnf} and \texttt{systemctl} idioms are well represented in pretraining corpora.

\section{Conclusion and Limitations}
\label{sec:conclusion}
\label{sec:limitations}

We introduced HardEval, a reusable benchmark of failed, automatable RHEL-8 DISA STIG controls, graded on a fixed 151-rule cross-model manifest (137 excluding SSH), with implementation-withheld natural-language prompts and OpenSCAP/OVAL fail-to-pass grading on real virtual machines. Across fourteen configurations, combined server-safe scores span 81.0\% (Claude Opus~4.7) down to 12.4\% (4-bit GLM-4-9B), with every frontier model above every open-weight one (Table~\ref{tab:hardeval-main}).

Per-domain reporting reveals what the aggregates hide: audit-rule construction is the sharpest discriminator between models; parameter count and release recency are weak predictors within tiers (Section~\ref{sec:variance}); four-bit quantization shows no detectable penalty for GLM-4-9B; and the recurring failures are value, mechanism, and audit-key mismatches rather than broken shell syntax. Two robustness checks support these findings: a GPT-4o-authored prompt set lowers scores for both solvers rather than favoring its author, and a zero-code port to Ubuntu~24.04 reproduces the pipeline on the 88 shared rules.

These results carry scope constraints: all graded tasks derive from RHEL~8 under the DISA STIG profile and cover a curated, automatable subset of controls; the Ubuntu port demonstrates methodological portability, but broader generalization to RHEL~9, Windows, or other profiles is unshown. Ten of the fourteen configurations were scored with a single greedy run, and the variance study excludes the closed models. SSH scores are conservative lower bounds because the harness grades over the same SSH channel the scripts harden, which is why that bucket is excluded from the headline metric (Section~\ref{sec:predictions}). Prompts were authored by a single model (addressed by the cross-author study, Section~\ref{sec:cross-author}), and training-data contamination from public STIG content cannot be ruled out. Finally, grading measures compliance-oracle satisfaction rather than security in the broad sense (Section~\ref{sec:hardeval-protocol}).

Taken together, these results establish scanner-verified compliance remediation as a first-class evaluation axis for language models. Future work includes extending HardEval to RHEL 9 and Windows STIG profiles and the open-source benchmark lets the community track this capability as models evolve.

\IEEEtriggeratref{8}
\bibliographystyle{IEEEtran}
\bibliography{ref_aaai_v2}

\end{document}